\documentclass{article}

    \usepackage[final]{neurips_2020}

\usepackage[utf8]{inputenc} % allow utf-8 input
\usepackage[T1]{fontenc}    % use 8-bit T1 fonts
\usepackage{hyperref}       % hyperlinks
\usepackage{url}            % simple URL typesetting
\usepackage{booktabs}       % professional-quality tables
\usepackage{amsfonts}       % blackboard math symbols
\usepackage{nicefrac}       % compact symbols for 1/2, etc.
\usepackage{microtype}      % microtypography
\usepackage{subcaption}
\usepackage{mathtools}
\usepackage[normalem]{ulem}
\useunder{\uline}{\ul}{}
\title{Optimizing One-pixel Black-box Adversarial Attacks}

\author{%
  Tianxun Zhou\\
  National University of Singapore\\
  \texttt{e0486478@u.nus.edu} \\
  \And
  Shubhankar Agrawal\\
  National University of Singapore\\
  \texttt{e0925482@u.nus.edu} \\
  \And
  Prateek Manocha\\
  National University of Singapore\\
  \texttt{e0674507@u.nus.edu} \\
}

\begin{document}

\maketitle

\begin{abstract}
The output of Deep Neural Networks (DNN) can be altered by a small perturbation of the input in a black box setting by making multiple calls to the DNN. However, the high computation and time required makes the existing approaches unusable. This work seeks to improve the One-pixel (few-pixel) black-box adversarial attacks to reduce the number of calls to the network under attack. The One-pixel attack uses a non-gradient optimization algorithm to find pixel-level perturbations under the constraint of a fixed number of pixels, which causes the network to predict the wrong label for a given image. We show through experimental results how the choice of the optimization algorithm and initial positions to search can reduce function calls and increase attack success significantly, making the attack more practical in real-world settings. 

\end{abstract}

\section{Introduction}
Deep Learning, a data-driven technology that can precisely model complex mathematical functions over large data sets, can successfully reach human-level performance for tasks such as image recognition. However, studies such as  \citep{szegedy2013intriguing, goodfellow2014explaining, kurakin2016adversarial, nguyen2015deep} show that these deep neural networks (DNN) models are vulnerable to delicately crafted images called "adversarial examples" that are generated by adding calculated artificial perturbation onto the raw image to make the DNN misclassify. \citep{szegedy2013intriguing} observed that the model predictions could be manipulated with minimal input perturbations, which appears unnoticeable to human eyes (maybe an image here). Discovered initially for image classification task \citep{szegedy2013intriguing} and now applicable to other computer vision tasks such as object detection \citep{tu2020physically, zhang2019towards} and semantic segmentation \citep{arnab2018robustness}  as well, the target of the generated adversarial examples can be summarised as to achieve misclassification while being unnoticeable and as similar to the original image as possible. 

While the key idea for creating adversarial images is to add the minimal amount of carefully crafted additive perturbation to achieve the goal of misclassification, the generated sample should also be as “similar” as possible to the original model input. This similarity is measured with metrics such as $L_0$, $L_1$ or $L_{\infty}$ norms. In real-world settings, where the attacker only has access to outputs of the model, black-box attacks have been proposed which do not require internal information of the model, such as gradients. Unfortunately, some of the previous attacks may not be useful in practical scenarios for adversarial attacks. Namely, the modifications might be excessive (i.e., the amount of modified pixels is relatively large) such that it may be perceptible to human eyes or that it requires making a very high number of calls to the DNN. Example of such includes StrAttack \citep{xu2018structured} which perturbs a large number of pixels to achieve the attack and One-pixel attack \citep{su2017onepixel} which achieve limited perturbation as measured by $L_0$-norm. However, it requires making a very high number of calls to the DNN, essentially making them unusable in the real world due to multiple factors such as time taken to create such examples or being flagged by the model provider due to excessive call being made to DNN. 

\begin{figure}
\centering
\begin{subfigure}{\textwidth}
   \includegraphics[width=\textwidth]{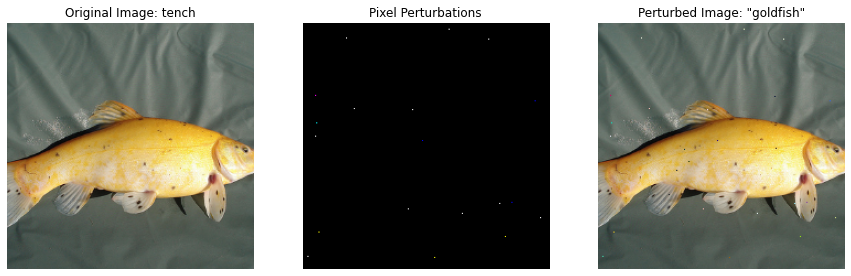}
\end{subfigure}
\begin{subfigure}{\textwidth}
   \includegraphics[width=1\linewidth]{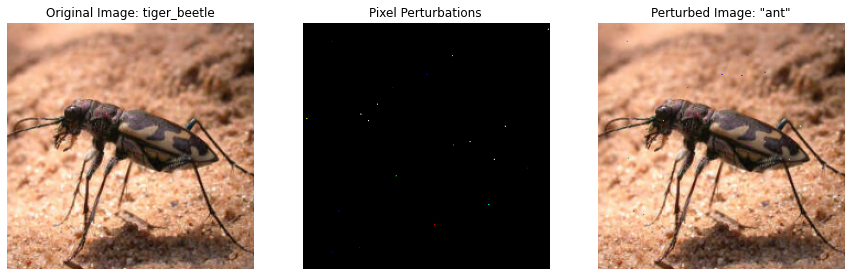}
\end{subfigure}
\begin{subfigure}{\textwidth}
   \includegraphics[width=1\linewidth]{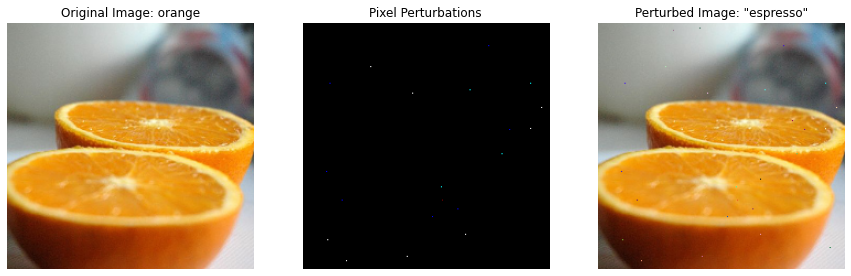}
\end{subfigure}
\begin{subfigure}{\textwidth}
   \includegraphics[width=1\linewidth]{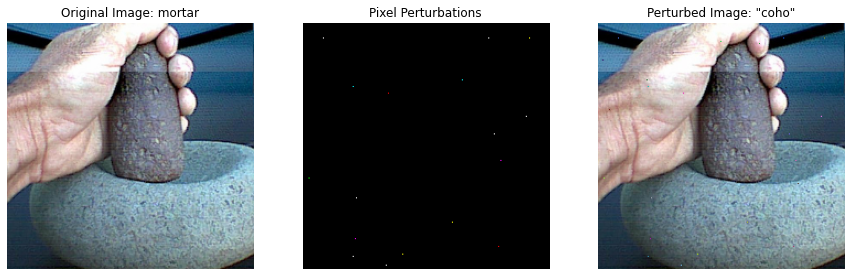}
\end{subfigure}
\caption{Adversarial Image Generated with Our Method}
\label{fig:method_demo}
\end{figure}

Thus, in order to facilitate the creation of an adversarial example that has a low $L_0$-norm, we introduce a method that is an improvement over the One-pixels attack that limits the number of perturbed pixels while significantly reducing the number of calls to DNN to obtain the perturbation. In particular, we use an alternative optimization algorithm (simulated annealing), initialized using a structural attack-derived perturbation mask that captures pixel-level spatial information from the input. Figure \ref{fig:method_demo} demonstrates an example of adversarial image generated using our method

Our  major contributions are summarized as below. We:
\begin{itemize}
  \item {demonstrate that simulated annealing as the optimization algorithm is able to achieve much higher attack success given a constraint on number of calls to the DNN}
  \item {propose an initialization scheme for the optimization process derived from structured attack, that further improves attack performance}
\end{itemize}

The rest of the article is organized as follows: In Section 2, we provide the details about the most related work in the field of adversarial attacks. In Section 3, we describe our novel approach and how it helps solve the problem of existing approaches. The experimentation results and ablation study is provided in Section 4, while the results obtained are discussed in Section 5. Finally, we conclude in Section 6.

\section{Related Works}

Literature Review encompassed reading through several papers to understand the different types of adversarial attacks on neural networks, followed by a deep dive into black-box adversarial attacks.

\paragraph{Black-box adversarial attacks.}
Black-box adversarial attacks refer to adversarial attacks that do not need internal knowledge, such as gradients, of the network being attacked. This is in contrast to white-box attacks such as gradient-based attacks, which compute the gradient of the loss with respect to the input image. Black-box attacks can generally be classified under three categories: transfer-based, score-based, and decision-based attacks.

Transfer-based attacks \citep{liu2017blackbox, cheng2019blackbox} uses a white-box model to generate adversarial examples to attack the black-box model. The white-box model uses a similar (but need not be identical) type of architecture as the black-box model, for e.g., varieties of Convolutional networks for image tasks. The assumption is that there is transferability between models trained on similar architecture and training processes. Score-based attacks \citep{ilyas2018blackbox, su2017onepixel, xu2018structured, chen2017blackbox} use the output probability scores by the model to formulate the attack. In score-based methods, different strategies such as gradient approximation and derivative-free optimization are used. Decision-based attacks use only the hard-label predictions by the model. Some decision-based attacks include boundary attack \citep{brendel2018blackbox} and optimization-based method \citep{cheng2018blackbox}.

Our method, which will be elaborated further in Section 3, falls under score-based attacks. It is an improvement upon the One-pixel attack and incorporates elements from structured attacks, both of which are score-based black-box adversarial attacks.

\paragraph{One-pixel attack.}

One-pixel or few pixels attacks proposed by \citep{su2017onepixel} is a method for generating adversarial perturbations on one or a small number of pixels. It uses differential evolution (DE) \citep{storn1997DE}, a genetic
optimization algorithm to find a perturbation of $d$ number of pixels. Each perturbation is defined as a tuple holding five elements: $(x,y)$ coordinates and the RGB value of the perturbation. The DE algorithm seeks to maximize the objective function of the adversarial loss of the model on an image given a set of perturbations.

\[
\underset{p(x)}{maximize} \;\;\; f_{adv}(x + p(x))
\]
\[
subject\;\; to \;\;\; \lVert p(x)\rVert \leq d
\]
where $f_{adv}$ is the adversarial loss of the network, for e.g. (1 - the softmax value of the true class), $x$ is the image, $p(x)$ is the perturbation, and $d$ is constraint set on number of pixels to perturb.

In the original paper and existing open-source implementations available, DE is used as the optimization algorithm. As a genetic algorithm, it does not need gradient information in the optimization process, and hence allowing the attack to be made in a black-box fashion.

\paragraph{Structured Adversarial Attack (StrAttack)}

The Structured Adversarial Attack proposed by \citep{xu2018structured} provides a novel way to approach adversarial attacks by aiming to understand the structure of the input image to generate relevant perturbations. The working of this method incorporates a sliding mask across the pixels of the input image to identify key structures. Group sparsity is then regularized and the loss function is optimized for the adversarial inputs over several iterations to produce results that can be interpreted. In addition, the StrAttack uses a alternating direction method of multipliers (ADMM) optimizer (first introduced by \citet{ghadimi2014optimal}) in comparison to conventional optimizers such as ADAM to solve its optimization problem.

\paragraph{Derivative-free optimization}

Derivative-free optimization methods are designed to optimize functions without requiring gradient information, which is useful for many real-world applications where function is unknown or where derivatives may not exist. For this reason, they can be useful tools when formulating a black-box adversarial attack on neural networks. As mentioned, differential evolution \citep{storn1997DE} used in One-pixel attack is a type of derivative-free optimization algorithm. There is generally no guarantee that optimizer can converge upon the global or even local optima within finite time. However, supported by many empirical studies, it has been found that in practice, they often can give good results.\citep{chopard2018optimization} Genetic/evolutionary algorithms, simulated annealing and swarm based algorithms are established and commonly used methods.

\section{Method}

As part of the methodology, augmentations to the One-pixel attack were performed by incorporating different methods summarized in this section.

\subsection{Using simulated annealing for optimization}
Inspired by the original One-pixel attack, but realizing that DE requires a large number of function evaluations to find the solution, alternative optimization algorithms were explored to replace DE. Simulated annealing (SA) \citep{kirkpatrick1983SA} is a well known and popular derivative-free optimization algorithm. It is inspired by annealing process in physics, where a material is heated up to a high temperature, and allowed to cool slowly. In nature, annealing allows materials to attain a lower energy configuration. Similarly, SA has a temperature parameter which starts off at a high value and decays slowly with each iteration. At each iteration, a neighbor position is chosen around the current solution candidate, and evaluated on the objective function.

\[
\Delta E = f(x^{'})-f(x)
\]
where $f(x)$ is the function value of the current solution candidate and $f(x^{'})$ is the function value of the newly chosen neighbor. Whether the new candidate is accepted to become the new solution is decided probabilistically, as a function of both $\Delta E$ and current temperature value. In simple SA, the probability of acceptance can be defined as such:

\[ P_a = \begin{cases}
1 \quad& if \;\; \Delta E \leq 0 \\
e^{-\Delta E/T} \quad& if \;\; \Delta E > 0\\
\end{cases}
\]

where $P_a$ is the probability of acceptance of the neighbor as the new solution candidate, and $T$ is temperature parameter. It is obvious that as temperature parameter starts high, it is more likely for the algorithm to select candidates that had higher objective function value. However as temperature decays, the probability decreases. In other words, the algorithm favors more exploration at the start of the process, gradually moving to exploitation as temperature parameter decays. In our method, we make use of a more advanced version of simulated annealing, Generalized simulated annealing proposed by \citep{tsallis1996SA}, which has more sophisticated mechanism for defining the temperature decay schedule, as well as the neighborhood of visit and acceptance criteria, as functions of temperature. 

The temperature decay schedule is defined as follows:
\[
T_{q_v}(t) = T_{q_v}(1) \frac{2^{q_v -1}-1}{(1+t)^{q_v -1}-1}
\]
where $q_v$ is a hyperparameter, the visiting parameter, and $t$ is the time, i.e. iteration. $T_{q_v}(1)$ is thus the initial temperature.

More details of the algorithm can be found in the paper by \citep{tsallis1996SA}, and scipy's documentation \citep{scipy2018SA}.

\subsection{Better initialization of optimization with image structure information}
Following the algorithms from the paper, a PyTorch implementation of the StrAttack is developed. As per the methodology by \citep{xu2018structured}, the following optimization problem is solved:
\[
\underset{\delta}{minimize} \;\;\; f(x_0 + \delta) + {\gamma}D(\delta) + {\tau}g(\delta)
\]
where $f$ indicates the loss function to create the perturbations, $D$ measures distortion indicating dissimilarity between original image and perturbed image, and $g$ symbolizes the l-norm of perturbation. The variables $\tau$ and $\gamma$ act as regularization parameters 

\begin{figure}[h]
  \centering
  \includegraphics[width=\textwidth]{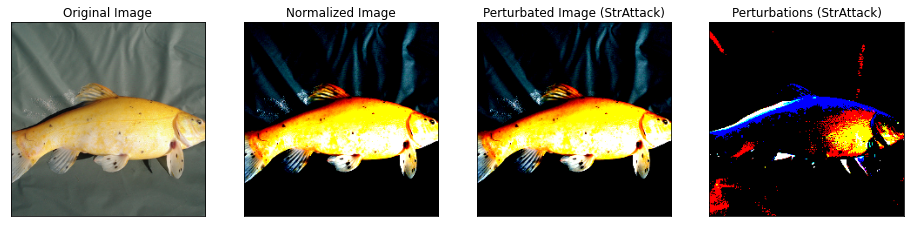}
  \caption{StrAttack Perturbations Generated}
  \label{fig:strattack_raw}
\end{figure}

Initial experiments are performed using StrAttack as the main adversarial attack to attain a working method of attack. An individual working StrAttack provides good results reducing the network accuracy. Images from a naive StrAttack implementation as in Figure \ref{fig:strattack_raw} show the perturbations generated being of small magnitude and indiscernible to the human eye severely reducing accuracy of the learning model. However perturbations over a number of pixels greater than the scope of our One-pixel Attack experiments are generated. 

\begin{figure}[h]
  \centering
  \includegraphics[width=\textwidth]{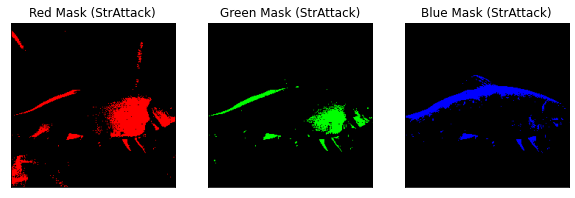}
  \caption{StrAttack Mask across RGB channels}
  \label{fig:strattack_mask}
\end{figure}

With a viable implementation of the StrAttack, the perturbations are extracted and using a constant threshold obtained via empirical methods, a binary mask over RGB levels is generated for each of the images as indicated in Figure \ref{fig:strattack_mask}.

This mask is then used as an initialization to the One-pixel attack to compare performance. Specifically, we choose $n$ number of pixels randomly where the binary mask is positive. The colors for those pixels are then defined to be the inverse of the values in the image. These pixels are used as the initial perturbations for the SA optimization algorithm to proceed upon.

\begin{figure}[h]
  \centering
  \includegraphics[scale=0.25]{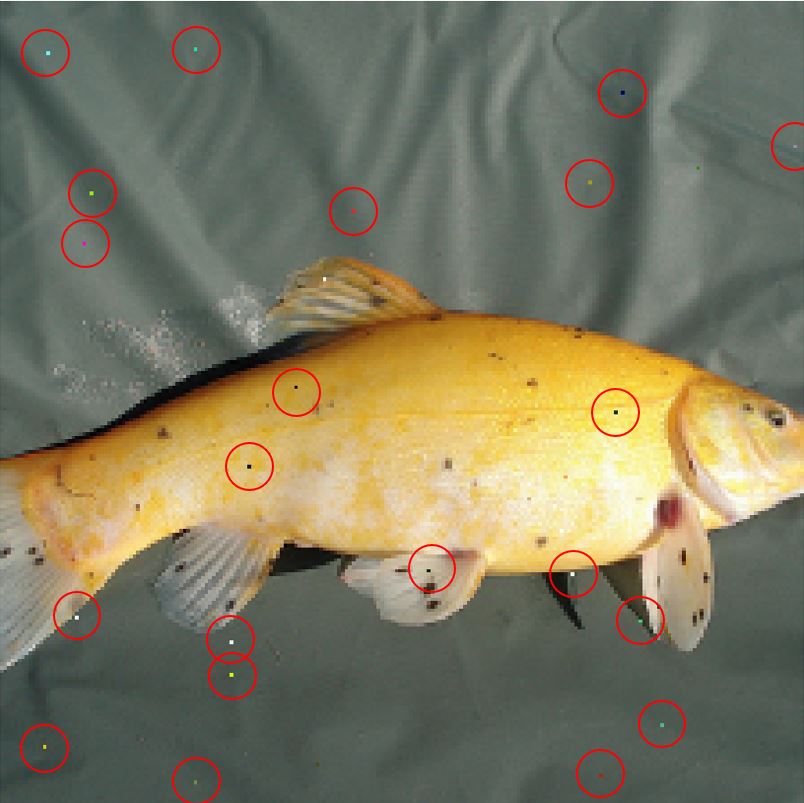}
  \caption{Image with few pixel perturbations found by our method}
  \label{fig:pixel_attack}
\end{figure}

Figure \ref{fig:pixel_attack} shows a perturbed image found by our few pixel attack method that successfully caused the model to predict an incorrect class. Pixel changes are circled in red. As seen, the few pixel changes were almost imperceptible to human subjects. 

\section{Experiments}

Having listed out the methods we aimed to work with, experiments were performed across those with standard configuration and parameters maintained throughout. The results of our experiments are covered in this section. Our code is made available to run in Google Colaboratory at \url{https://drive.google.com/drive/folders/1elews75KB49_hKW6WKXgpq-zlf46piWO?usp=sharing}

\subsection{Experiment set-up}
The attack is performed on ResNet-18 pre-trained on Imagenet. For the attack, 100 images were chosen, 1 from each of the 1st 100 classes of Imagenet testset. ResNet-18 pre-trained model is provided by Pytorch. We benchmark our attack against the original One-pixel attack implemented in Torchattacks \citep{kim2020attacks}. 

The attack is a un-targeted attack, and we allow perturbation of 25 pixels. We set a limit on maximum number of 15000 function calls. If the algorithm is unable to find a solution within the maximum function call limit, then the attack is deemed as unsuccessful. After running on the same 100 test images, we compare the average number of function calls per successful attack, as well as the overall attack success rate.

\subsection{Comparison with original One-pixel attack}
The results show that our method is able to achieve significantly higher attack success rate than the original OP attack, and the average number of function calls is also lower. The average number of function calls made for successful attacks however, is lower in the original OP attack. This is due to the fact that there were lesser successful attacks by the original OP attack. The successful ones were likely easier to attack in the first place, hence resulting in faster attack.

\resizebox{\textwidth}{!}{%
\begin{tabular}{@{}lcccl@{}}
\toprule
\textbf{Attack}     & \textbf{Network Accuracy (\%)} & \textbf{Mean Calls for Successful Attacks} & \textbf{Mean Calls for All Attacks} &  \\ \midrule
None                & 83                             & N/A                                        & N/A                                 &  \\
Original OP Attack  & 72                             & \textbf{829}                              & 11032                               &  \\
\textbf{Our Method} & \textbf{13}                    & 2189                                       & \textbf{3598}                       &  \\ \bottomrule
\end{tabular}%
}

\subsection{Ablation study}
In this section, we investigate the effect of only simulated annealing without structured attack initialization. 

\resizebox{\textwidth}{!}{%
\begin{tabular}{@{}lcccl@{}}
\toprule
\textbf{Attack}          & \textbf{Network Accuracy (\%)} & \textbf{Mean Calls for Successful Attacks} & \textbf{Mean Calls for All Attacks} &  \\ \midrule
Random Initialization    & 13                             & \textbf{1855}                              & 4096                                &  \\
StrAttack Initialization & \textbf{11}                    & 2189                                       & \textbf{3598}                       &  \\ \bottomrule
\end{tabular}%
}

The results of the ablation study shows that StrAttack initialization provides a slight improvement of 2\% in attack success, and hence a lower mean calls for all attacks. However, the mean calls for successful attacks is higher for StrAttack initialization. Note that to generate the StrAttack mask, 5 calls to the network was made. These have been already accounted for in the results presented.

\section{Discussions}
\paragraph{Effectiveness of simulated annealing compared to differential evolution.}
It is demonstrated through the experiments that SA is highly effective compared to DE on performing black-box attack on neural networks. We hypothesize a few possible reasons for this. 

The first is that SA may be better suited for discrete problems \citep{chopard2018optimization}, while DE was originally designed for continuous problems \citep{storn1997DE}. The model loss as a function of pixel perturbation is a discrete problem as pixel position and RGB values all take on discrete values, and hence SA may perform better.

The second reason is that DE is a population based algorithm where during each iteration, a population of candidates are generated and evaluated. This setup inherently favors more exploration by sampling sampling more in the search space. However when examining the mask generated by StrAttack, it can be seen that for many images, the pixels that have greater impact on model prediction is often concentrated at particular areas of the image. Exploring solutions candidates spread out evenly over the entire image may not be very efficient. Hence population based algorithms such as DE would perform a large number of function calls that explores the search space but ultimately may not help in finding the solution.

\paragraph{Use of image structure information for initialization of optimization}
The use of mask generated by StrAttack for initialization shows a slight improvement on attack success. In derivative-free optimization research, it is well studied that initialization plays a very important role in the performance of these algorithms \citep{li2020optimization}. The high degree of interpretability provided by the StrAttack into the composition of the image is leveraged to improve results in this study. StrAttack generated mask provides prior knowledge that serves as an intuitive way to initialize the optimization. 

\section{Conclusion}
This work targets to optimize the creation process of adversarial examples by limiting the number of pixels perturbed and reducing the number of calls made to DNN. Unlike the previous approaches, our work demonstrates that simulated annealing as the optimization algorithm improves over the One-pixel attack by significantly reducing the number of calls to DNN. In order to further optimize the performance, we leverage on Structured Attacks to derive the pixel-level structural information for use as an initialization scheme. We perform experiments using the Imagenet dataset showing the effectiveness of our proposed approach and its better generation speed for adversarial examples.

\section{Future Work}
Given the significant difference between SA and DE in performance as measured by attack success under constraint of function evaluations to the DNN, future work may be directed at understanding how the adversarial loss landscape is like with respect to pixel perturbations. The characteristics of the adversarial loss landscape may be provide explanations on why certain optimization algorithms are more suited to the task, and will lead to more theoretical insights to guide design of better attacks.

Another direction of work can be directed at devising more sophisticated ways of using structured attack information. Using the information for optimization initialization is just one method which was explored in this paper. Other methods may provide better results.

% \section*{References}

\bibliography{references}

\end{document}